\documentclass{PoS}

\usepackage{xcolor}
\usepackage{subfigure}
\usepackage{mcite}

\title{Thermal noise in complex systems}

\ShortTitle{Thermal noise in complex systems}

\author{Rene Glaser, Klara T. Knupfer, Lukas Maczewsky, Max M{\"a}usezahl, Ronny Nawrodt\\
        Friedrich-Schiller-Universit{\"a}t Jena, Institut f{\"u}r Festk{\"o}rperphysik, Helmholtzweg 5, D-07743 Jena, Germany\\
        }
				
\author{Garrett D. Cole\\
        Crystalline Mirror Solutions LLC and GmbH, Santa Barbara, CA, 93101 USA and 1060 Vienna, Austria
\\
        }

\author{Johannes Dickmann, \speaker{Stefanie Kroker} \\
        Physikalisch-Technische Bundesanstalt, Bundesallee 100, D-38116 Braunschweig, Germany \\ Technische Universit\"at Braunschweig, LENA Laboratory for Emerging Nanometrology, Am~Langen~Kamp~6b, D-38106 Braunschweig, Germany\\
        E-mail: \email{stefanie.kroker@ptb.de}}

\abstract{We present a method to calculate the power spectral density of Brownian noise in complex optomechanical systems using Levin's approach of virtual pressure and present first mechanical loss measurements for high-purity GaAs over a wide temperature range from 7\,K to 250\,K. The loss reveals three Debye loss peaks. Each peak corresponds to an Arrhenius-like relaxation process with activation energies of 17.9\,meV, 65.4\,meV and 123\,meV respectively. Additional light induced damping was observed for photon energies below and above the fundamental gap of GaAs in contrast to observations by Okamoto \textit{et al}.}

\FullConference{GRAvitational-waves Science\&technology Symposium - GRASS2018\\
		1-2 March 2018\\
		Palazzo Moroni, Padova (Italy)}

\begin{document}

\section{Introduction}

Optomechanical high-precision instruments ranging from microscale devices such as sensors \cite{Balram2014} up to km-size interferometers \cite{LIGO, Virgo,ET, KAGRA} serve as key ingredients for the newly born field of gravitational wave astronomy. Their performance is limited by several noise sources. A significant contribution arises from Brownian thermal noise which is linked to the mechanical loss of the construction material via the fluctuation-dissipation theorem \cite{Callen1951}. Crystalline materials such as silicon, sapphire or gallium arsenide (GaAs) with low mechanical losses are promising for low thermal noise operation \cite{Balram2014,ET, KAGRA}. GaAs is of particular interest for optomechanical devices due to its large photoelastic coefficient and well established fabrication techniques \cite{Balram2014}. Furthermore, GaAs-based crystalline coatings have demonstrated extremely low thermal noise performance and high reflectivity \cite{Cole2013, Cole2016}. In many cases a combination of different materials in different shapes is needed to assemble the optomechanical devices forming a very complex system. 

In this contribution we describe a path to how to calculate thermal noise in such complex devices. Additionally, we present first investigations of the mechanical loss of GaAs-cantilevers in a wide temperature range from 7\,K to 250\,K. This parameter significantly affects the level of Brownian thermal noise in a system.

\section{Brownian thermal noise in optomechanical systems}

The Brownian thermal noise of optomechanical systems can be calculated using Levin's approach \cite{Levin1998}. On plain surfaces where all light is reflected directly at the surface, a virtual pressure with the same spatial distribution as the laser beam intensity (e.g. a Gaussian distribution) is used to determine the mechanical strain distribution and thus the deformation energy density $\epsilon(\vec{r})$ within the sample. This deformation energy distribution then serves to compute the energy dissipated in the system at a given frequency $f$: 

\begin{equation}
\label{Eq:Wdiss}
W_\mathrm{diss}(f,T)=2\pi f\int{\epsilon(\vec{r})\phi(\vec{r},T)} \mathrm{dV},
\end{equation}

\noindent where $\phi(\vec{r})$ is the spatial distribution of the mechanical loss under the assumption of structural damping \cite{Nowick1972}. The integration has to be performed over the entire sample volume. From the dissipated energy, the power spectral density of Brownian thermal noise $S_\mathrm{z}(\omega, T)$ is computed by:

\begin{equation}
\label{Eq:FDT}
S_\mathrm{z}(\omega, T)=\frac{2k_\mathrm{B}T}{\pi^2f^2}\frac{W_\mathrm{diss}}{F_0^2}. 
\end{equation}

\noindent  Here, $k_\mathrm{B}$ represent Boltzmann's constant, $T$ the temperature, and $F_0$ the virtual force that was initially applied to the surface of the test mass. Recently, it was shown that the applied virtual pressure has to be weighted with the radiation pressure difference at the surface. It is given by the difference of the respective components of radiation pressure at the surface which is determined by Maxwell's stress tensor \cite{Kroker2017, Tugolukov2017}. Reducing thermal noise requires cryogenic temperatures, materials with low mechanical losses and/or a deliberate distribution of the mechanical losses. If materials are needed that have large mechanical losses they have to be placed in locations where the strain energy, i.e. the radiation pressure and hence the electromagnetic field, is small. Maxwell's stress tensor enables the rigorous computation of thermal noise and thus  the holistic design of high-reflectivity low-noise coating stacks \cite{Cole2013,Steinlechner2015,Yam2015,Lovelace2017} and surface structures \cite{Friedrich2011,Kroker2013, Dickmann2017}. 

The mechanical loss $\phi$ of a sample is governed by complex solid state dynamics \cite{Nowick1972} which are up to now not fully understood. Interfaces \cite{Nawrodt2013}, impurities or temperature strongly affect the mechanical loss and can cause variations of this parameter by orders of magnitude and change the thermal noise of the system. In the following section we present loss measurements of high-purity GaAs flexures. They show a dependence of the mechanical loss on illumination.

\section{Loss measurement of GaAs flexures}

The mechanical loss $\phi$ in single-crystal GaAs samples has been studied by means of a cantilever ring-down technique \cite{Nawrodt2013,Reid2006}. GaAs wafers with (100) orientation have been cut in $[110]$ direction to typically 50\,mm long and 5\,mm wide beams. These were thinned to less than 0.1\,mm using a wet chemical etchant consisting of $\mathrm{NH}_4\mathrm{OH}$, $\mathrm H_2 \mathrm O_2$ and deionized water. The samples have been excited to resonant vibrations by means of a high-voltage driving plate. The resonant vibration has been read out using an optical lever technique utilizing a quadrant photo-diode. Previous investigations have shown that GaAs exhibits a light induced damping mechanism \cite{Okamoto2011a, Okamoto2011b} when illuminated with photons having a larger energy than the fundamental gap of the sample. Here, two different wavelengths have been used for the optical lever setup, a red laser with $650\,\mathrm{nm}$ in vacuum and an IR-laser with $1550\,\mathrm{nm}$. These wavelengths correspond to about $1.9\,\mathrm{eV}$ and $0.8\,\mathrm{eV}$ respectively and therefore to energies above and below the fundamental gap at around $1.5\,\mathrm{eV}$ (corresponding to a wavelength of $\approx830\,\mathrm{nm}$) \cite{Grilli1992}.

\begin{figure}
	\centering
	\subfigure[]{\includegraphics[width=0.49\linewidth]{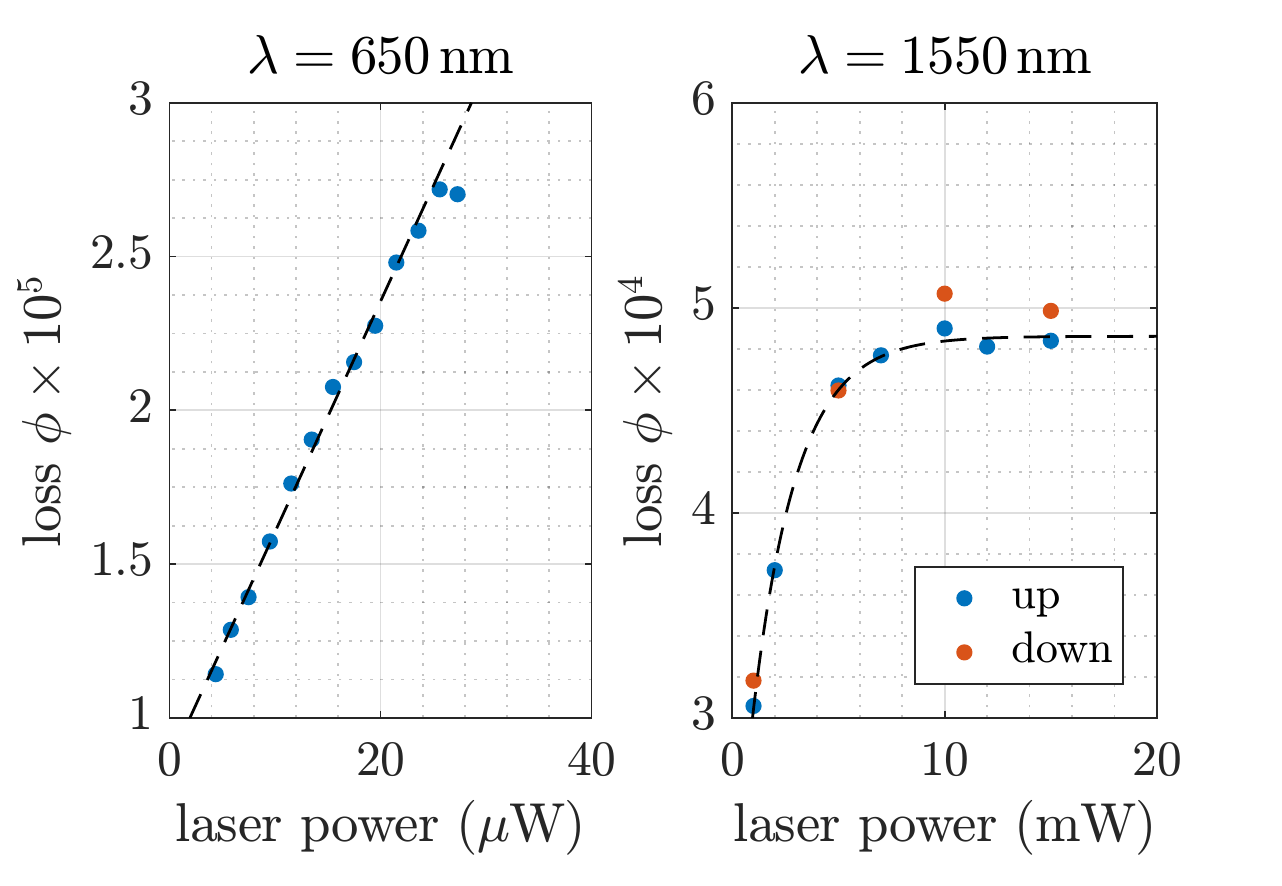}\label{fig:80K}}
	\subfigure[]{\includegraphics[width=0.49\linewidth]{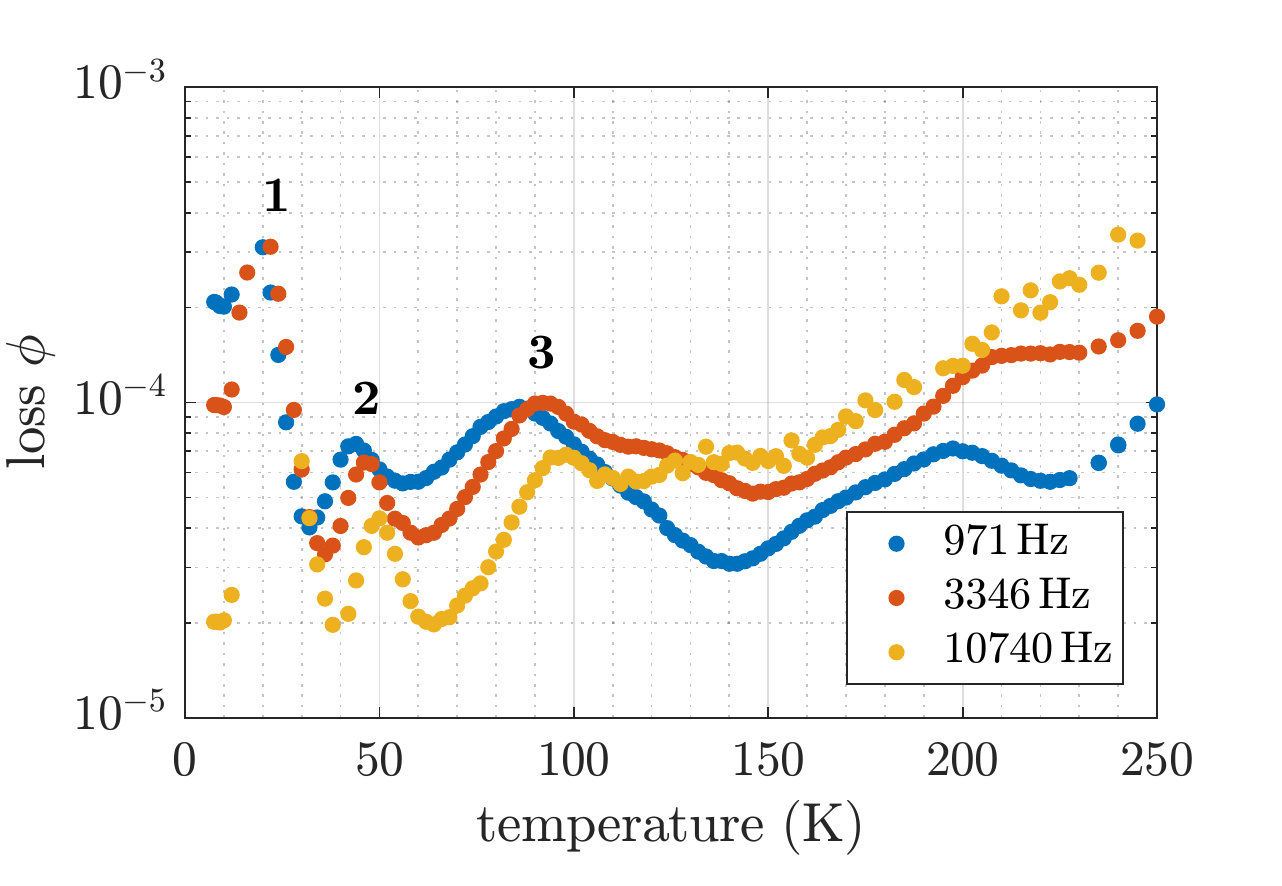}\label{fig:LHeup}}
	\caption{(a) For photon energies higher as well as lower than the fundamental gap illumination affects the mechanical loss at 80\,K of the GaAs cantilevers. The investigations have been done with vibrational modes at around 3\,kHz. The dashed lines shall guide the eye. (b) Three loss peaks can clearly be seen over a temperature range between 7\,K and 100\,K as shown here for example for three transverse vibration modes.}
\end{figure}

For low illumination power a linear increase of the mechanical loss with the illumination power is observed (see figure\,\ref{fig:80K}). At higher illumination power a saturation can be observed. For both cases, photon energies larger and smaller than the fundamental gap of GaAs, photo-induced damping is observed which is in contrast to the previous studies in \cite{Okamoto2011a, Okamoto2011b}. The observed photo-induced damping leads to a hysteric behavior of the temperature dependent mechanical loss and cannot be caused by absorption induced heating effects. In order to gain insight into the complex solid state dynamics in GaAs the mechanical loss has been studied in a wide temperature range from 7\,K to 250\,K. The measurement has been done using a 1550\,nm laser with a constant laser power of 1\,mW. Figure\,\ref{fig:LHeup} shows three loss peaks at about 20\,K, 50\,K and 90\,K. These loss peaks are the so called Debye peaks assuming an Arrhenius-like process.

\begin{figure}
	\centering
	\includegraphics[width=0.5\linewidth]{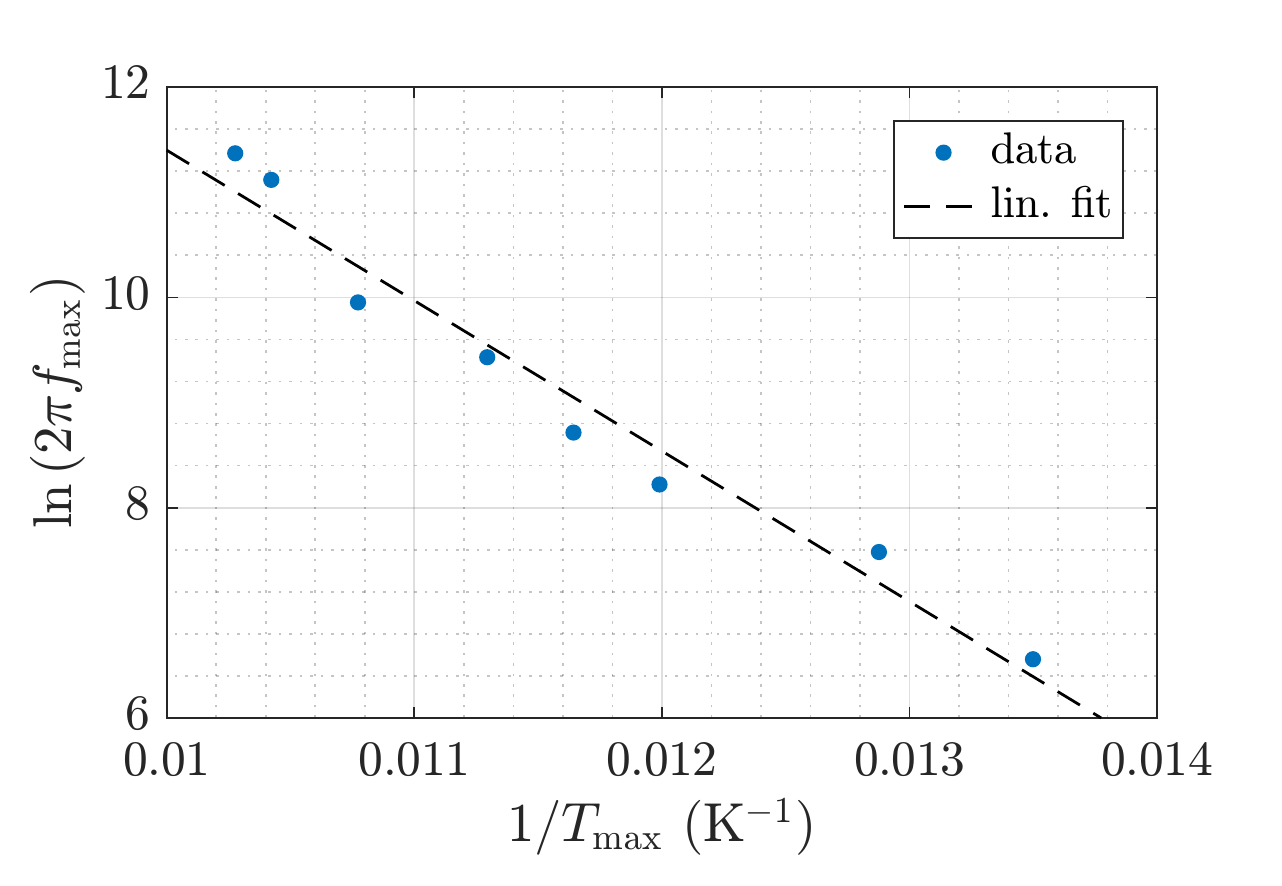}\label{fig:peak3}	\caption{Arrhenius plot for the loss measurement data of the third peak in figure\,\ref{fig:LHeup}, using all measured vibration modes. For each mode and peak, peak temperature $T_\mathrm{max}$ and peak frequency $f_\mathrm{max}$ were determined. The slope and the intercept of the linear fit correspond to an activation energy of 123\,meV and a relaxation constant of 6.9\,ps.}
\end{figure}

Debye peaks are usually a hint for relaxation processes in the solid that can be described by an activation energy and a relaxation time constant. Figure\,\ref{fig:peak3} presents the so-called Arrhenius plot for peak 3 in figure \ref{fig:LHeup}. The temperature $T_\mathrm{max}$ corresponds to the temperature of the observed maximum of the mechanical loss. This temperature was determined for different vibrational modes having a frequency $f_\mathrm{max}$. The corresponding activation energies to the three peaks in figure\,\ref{fig:LHeup} are ($18\pm1$)\,meV, ($65\pm1$)\,meV and ($123\pm10$)\,meV respectively, the relaxation constants are 2.8\,ns, 2.6\,ps and 6.9\,ps respectively. These parameters are currently being explored to better understand the origin of the mechanical loss in these samples.

\section{Conclusion}
The computation of Brownian thermal noise in complex systems for the first time allows the holistic design of low-noise optomechanical sensors with optimized sensitivity. Light induced damping in GaAs has been observed for photon energies above and below the fundamental gap. The mechanical loss has been investigated for a wide temperature range between 7\,K and 250\,K and three Debye peaks were observed revealing Arrhenius-like relaxation processes within the material.

\section{Acknowledgement}

R.G. acknowledges the support of his PhD by the local government of Thuringia.

\end{document}